\documentclass[journal]{IEEEtran}
\makeatletter
\def\endthebibliography{%
  \def\@noitemerr{\@latex@warning{Empty `thebibliography' environment}}%
  \endlist
}
\makeatother

\ifCLASSINFOpdf
	\usepackage[pdftex]{graphicx}
	\graphicspath{{./Images/}}
\else
	\usepackage[dvips]{graphicx}
	\graphicspath{{./Images/}}
\fi

\usepackage{amsmath}
\usepackage{amssymb}
\usepackage{amsfonts}
\usepackage{wasysym}
\usepackage{mathrsfs}
\usepackage{latexsym}
\usepackage{xfrac}
\usepackage{xcolor}
\usepackage{tikz}
\usepackage{booktabs}
\usepackage{multirow}

\usepackage{array}
\usepackage{enumerate}
\usepackage[para]{threeparttable}
\newcolumntype{P}[1]{>{\centering\arraybackslash}p{#1}}
\usepackage{longtable}

\usepackage{xr}

\newcommand{\suml}{\sum\limits}
            
\definecolor{myMATLABblue}{RGB}{0, 110, 191}
\definecolor{myDarkRed}{RGB}{219, 48, 0}
\definecolor{myFlashRed}{RGB}{255, 0, 0}

\newcommand{\colorblue}{\textcolor{myMATLABblue}}
\newcommand{\colordarkred}{\textcolor{myDarkRed}}

\begin{document}

\title{IMAGINE: An 8-to-1b 22nm FD-SOI Compute-In-Memory CNN Accelerator With an End-to-End Analog Charge-Based 0.15-8POPS/W Macro Featuring Distribution-Aware Data Reshaping}
%
%
%

\author{Adrian~Kneip,~\IEEEmembership{Member,~IEEE,}
		Martin~Lefebvre,~\IEEEmembership{Member,~IEEE,}
		Pol~Maistriaux,~\IEEEmembership{Graduate Student Member,~IEEE,}
        and~David~Bol,~\IEEEmembership{Senior Member,~IEEE}%
\thanks{This work was supported by the \textit{Fonds National de la Recherche Scientifique} (FNRS), Belgium, under CDR Grant J.0014.20 and A. Kneip’s Research Fellowship. A. Kneip and M. Lefebvre are with the EEMCS Department, Delft University of Technology, Delft, The Netherlands. P. Maistriaux and D. Bol are with the ICTEAM Institute, UCLouvain, Belgium. (corresponding author: a.kneip@tudelft.nl).}
}

\maketitle


\begin{abstract}

Charge-domain compute-in-memory (CIM) SRAMs have recently become an enticing compromise between computing efficiency and accuracy to process sub-8b convolutional neural networks (CNNs) at the edge. Yet, they commonly make use of a fixed dot-product (DP) voltage swing, which leads to a loss in effective ADC bits due to data-dependent clipping or truncation effects that waste precious conversion energy and computing accuracy. To overcome this, we present IMAGINE, a workload-adaptive \mbox{1-to-8b} CIM-CNN accelerator in 22nm FD-SOI. It introduces a 1152$\times$256 end-to-end charge-based macro with a multi-bit DP based on an input-serial, weight-parallel accumulation that avoids power-hungry DACs. An adaptive swing is achieved by combining a channel-wise DP array split with a linear in-ADC implementation of analog batch-normalization (ABN), obtaining a distribution-aware data reshaping. Critical design constraints are relaxed by including the post-silicon equivalent noise within a CIM-aware CNN training framework. Measurement results showcase an 8b system-level energy efficiency of 40TOPS/W at 0.3/0.6V, with competitive accuracies on MNIST and CIFAR-10. Moreover, the peak energy and area efficiencies of the 187kB/mm$^2$ macro respectively reach up to 0.15-8POPS/W and 2.6-154TOPS/mm$^2$, scaling with the 8-to-1b computing precision. These results exceed previous charge-based designs by 3-to-5$\times$ while being the first work to provide linear in-memory rescaling.

\vspace{0.15cm}

\end{abstract}
\vspace{-0.5cm}

%
\IEEEpeerreviewmaketitle
\begin{figure}[!t]
	\centering
	\includegraphics[width=0.5\textwidth]{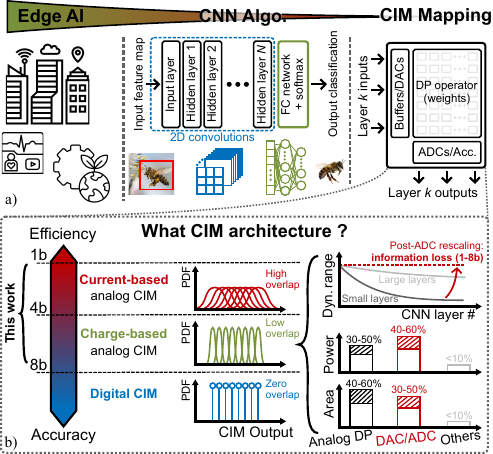}
	\caption{a) Top-down overview of mapping edge AI applications onto compute-in-memory (CIM) hardware for high-efficiency edge CNN processing. b) Precision scope of existing CIM architectures and illustration of the challenges faced by charge-based ones.}
	\label{Fig_1}
	\vspace{-0.4cm}
\end{figure}

\section{Introduction}

\IEEEPARstart{A}{s} of today, the fast-growing deployment of ever-more complex AI tasks in embedded systems has pushed conventional edge devices to their limit. Targeting a variety of biomedical, industrial or environmental applications, powerful AI algorithms such as convolutional neural networks (CNNs) require dedicated edge nodes that provide extreme levels of energy efficiency in order to save battery lifetime and avoid frequent replacement \cite{Shi_2016, Lauridsen_2018}. In that regard, compute-in-memory (CIM) architectures \cite{jintao_zhang_machine-learning_2016,Kang_2018} have rapidly gained attention for their ability to efficiently perform massively-parallel dot-product (DP) operations while bypassing the von-Neumann bottleneck, making them suitable candidates to accelerate CNNs at the edge, as depicted in Fig. \ref{Fig_1}(a). Nowadays, the current landscape of CIM implementations undergoes different trade-offs between computing efficiency and accuracy. On the one hand, analog CIM-SRAMs based on current-domain DP operators yield outstanding computing efficiency \cite{Sinangil_2021,Kneip_2023}, but are also highly sensitive to analog impairments such as nonlinearity, process variations and transistors mismatch, which hinder their computing accuracy \cite{jintao_zhang_machine-learning_2016,Kneip_2021}. While part of these non-idealities can be compensated during the off-line training of the CNN \cite{Kneip_2023}, their actual computing precision usually remains below 4b, restricting their applicative scope. On the other hand, digital CIM macros can support high-precision targets with near-golden accuracy, but waste efficiency due to the overhead of distributed adder trees \cite{Fujiwara_2022,Wang_ISSCC_2022} compared to their analog counterpart. In between, analog charge-based CIM-SRAMs offer an interesting compromise, relying on variation-insensitive metal-oxyde-metal (MoM) capacitances to perform analog DPs with a moderate area overhead \cite{Valavi_2019,Jia_2021}. With recent progress in quantization-aware training relaxing the precision requirements below 8b for many edge applications \cite{Han_2016, Baskin_2021}, charge-based architectures emerge as appealing candidates for the deployment of versatile CNN workloads at the edge.

Nonetheless, charge-based architectures come with their own challenges, qualitatively illustrated in Fig. \ref{Fig_1}(b). First, previous works that rely on charge-injection \cite{Lee_2021,Wang_VLSI_2022} utilize a fixed analog DP voltage swing, bound to the maximum array size. Yet, this approach reduces the effective number of available ADC bits when mapping small- or medium-sized CNN layers that do not require full macro utilization. Furthermore, combining a conventional full-scale ADC with such a fixed voltage swing introduces either clipping or truncation of the ADC output depending on the inbound DP data distribution. Altogether, these issues lead to wasted ADC area and energy, as well as to a loss of information that cannot be recovered by post-ADC rescaling. Finally, supporting a scalable 1-to-8b computing precision puts harsh constraints on DAC and ADC designs, consuming a significant fraction of the total energy and area of the macro and jeopardizing its flexibility towards small workloads. Although fully-serial implementations have been proposed to ease on that side \cite{Jia_2020}, they suffer from a costly 8b ADC conversion per input precision bit. 

In this work, we present IMAGINE, a massively-parallel CIM-CNN accelerator embedding a 1-to-8b \textbf{I}n-\textbf{M}emory-computing SRAM with end-to-end \textbf{A}nalog char\textbf{G}e-domain comput\textbf{I}ng and distributio\textbf{N}-aware data r\textbf{E}shaping. The featured macro combines a channel-wise swing-adaptive DP operator with an in-ADC multi-bit analog batch-normalization (ABN) function. Flexible bit-precision scaling is enabled by a novel input-serial, weight-parallel post-DP accumulation scheme. The CIM-SRAM macro is co-designed with hardware-aware CNN training to provide resilience against residual nonlinearity and variability, enabling approximate LSB computing at the 8b precision to relax the ADC design constraints. IMAGINE has been implemented in 22nm FD-SOI within the CERBERUS chip, with measurement results showcasing peak 8b energy efficiencies of 150 and 40TOPS/W for the standalone macro and the entire accelerator, respectively. The CIM-SRAM also reaches competitive throughput with a high 187kB/mm$^2$ density, improving overall performance metrics over the state of the art, while being the first to propose a linear in-memory gain rescaling. 

The rest of this paper is outlined as follows. Section II zooms in on the working principle and related challenges of configurable charge-based CIM designs. Then, Section III presents the CIM-SRAM macro architecture, while Section IV covers the accelerator's digital dataflow. Finally, Section V discusses measurement results.


\section{Basics and Challenges of Charge-Based CIM}

\begin{figure}[!t]
	\centering
	\includegraphics[width=0.5\textwidth]{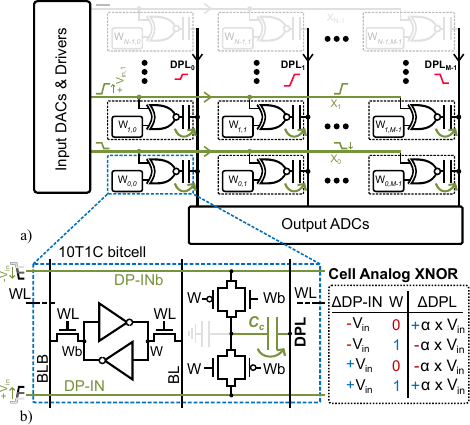}
	\caption{a) Simplified view of charge-based CIM-SRAM architectures, which accumulate the local results of analog XNORs from each b) 10T1C bitcell by means of charge injection through their computing capacitance $C_c$ on the column-shared dot-product line (DPL).}
	\label{Fig_2}
	\vspace{-0.4cm}
\end{figure}

In order to deal with the high inherent nonlinearity and variability of current-domain architectures, charge-based designs leverage variation-insensitive MoM capacitors to complete the analog DP operation with high accuracy. A typical architecture of charge-based CIM-SRAM relying on capacitive coupling is depicted in Fig. \ref{Fig_2}(a), focusing on the array of DP operators. After the DAC conversion, non-zero inputs $\mathrm{X}$ are propagated horizontally on the differential input lines DP-IN and DP-INb. Then, an analog XNOR operation takes place within each 10T1C bitcell, as represented in Fig. \ref{Fig_2}(b): depending on the value of the stored binary weight $\mathrm{W}$, acting as a +1/-1 factor, the analog XNOR outputs of each cell are accumulated on their shared dot-product line (DPL) by charge injection through the coupling MoM capacitance $C_c$, such that

\vspace{-0.1cm}

\begin{equation}
	V_{DPL,j} = V_{DD}/2 + \alpha \suml_{i = 0}^{N_{on}-1} \Delta{V_{in,i}} \, \mathrm{W}_{i,j}
	\label{Eq_1}
\end{equation}

\noindent where

\vspace{-0.1cm}

\begin{equation}
	 \Delta{V_{in,i}} = \dfrac{1}{2^{r_{in}}} \suml_{k = 0}^{r_{in}-1}{(-1)^{1-\mathrm{X}_i[k]} \, 2^k \, V_{DD}}.
	\label{Eq_2}
\end{equation}

\noindent $r_{in}$ is the bitwise input precision, $N_{on}$ the number of activated rows and $\alpha$ the charge-injection attenuation factor given by $\alpha = 1/N_{rows}$, ignoring parasitics and other DPL loads. While Eq. (\ref{Eq_1}) demonstrates an excellent linearity, it also points out that covering the full $V_{DD}$ voltage swing of the DPL is only possible at maximum array utilization, and decreases linearly with $N_{on}$. 

\begin{figure}[!t]
	\centering
	\includegraphics[width=0.5\textwidth]{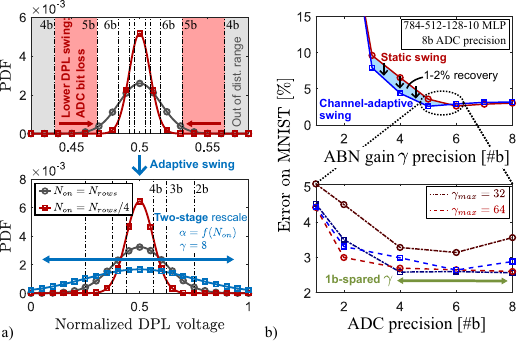}
	\caption{a) Considering 8b ADCs, narrow normal distribution of DPL voltages in charge-based CIM-SRAMs lead to multiple wasted precision bits during the conversion. Voltage swing reduction with $N_{on} < N_{rows}$ further reduces this effective number of ADC bits. Providing (i) channel-adaptive DPL range compensation and (ii) pre-ADC ABN rescaling help solve these issues. b) Test error on the MNIST dataset for a 784-512-128-10 MLP with various ABN gain and ADC precisions. Providing a channel-adaptive swing on top of ABN rescaling trades accuracy recovery for $\gamma$ precision.}
		\label{Fig_3}
	\vspace{-0.4cm}
\end{figure}

The inability to rescale the DPL swing penalizes the mapping of layers with narrow DP distributions, as well as that of layers not utilizing the entire input span (e.g., early layers in CNNs), as they fail to fully utilize the available ADC dynamic range. Taking an example in Fig. \ref{Fig_3}(a), a CNN layer with a zero-centred DP distribution that uses all (resp. 1/4th) of the CIM-SRAM inputs sees an effective ADC precision reduced by 2b (resp. 3b). This can result in a significant accuracy loss due to the network's inability to learn proper scaling factors, as observed in \cite{Kneip_2023} for ADCs below 6b. Moreover, this dynamic reduction in the effective number of ADC bits leads to wasted ADC power and area, reaching up to more than 75\% at 8b, thereby jeopardizing the efficiency of the analog computation.

In light of these challenges, it becomes paramount to be able to adjust the DPL swing based on the input dimension and the DP distribution of the target layer. Such rescaling can be obtained in two steps. First, by configuring the analog DP array to match the target input dimension, making $\alpha$ a function of $N_{on}$ so as to avoid wasting voltage swing and energy. Second, by providing data reshaping abilities prior to the ADC conversion, in order to linearly rescale and bias the DP distribution. This second part can be learned by the network, and thereby merged into a pre-ADC analog batch-normalization (ABN) with gain $\gamma$ and offset $\beta$:

\begin{equation}
	V_{ABN,j} = \gamma \; V_{DPL,j} + V_{\beta,j}.
\end{equation}

\noindent By modeling both stages during the CNN training in Fig. \ref{Fig_3}(b), we demonstrate that the supporting a channel-wise adaptive swing on top of the ABN rescaling saves 1b of $\gamma$ precision, simplifying the implementation of the ABN gain. Still, the design of these channel-adaptive and distribution-shaping stages has to be carefully carried out to minimize any downside effect on the linearity and variability of the CIM-SRAM operation. In particular, the linearity of the ABN implementation is key to map complex workloads, as opposed to the nonlinear ABN design in \cite{Kneip_2023} which is limited to MNIST-type problems. We thus address these concerns in what follows.



\section{Proposed CIM-SRAM Architecture}

\begin{figure}[!t]
	\centering
	\includegraphics[width=0.5\textwidth]{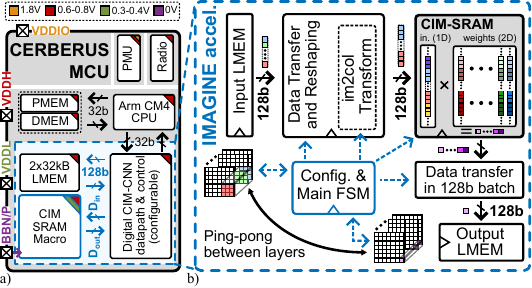}
	\caption{a) Block diagram of the CERBERUS micro-controller, zooming on b) the IMAGINE mixed-signal CIM-CNN accelerator.}
	\label{Fig_4}
	\vspace{-0.4cm}
\end{figure}

The proposed IMAGINE accelerator supports the 1-to-8b mapping of CNN layers with various dimensions, thereby offering flexibility on top of high efficiency. This CIM-CNN accelerator is embedded within the 22nm FD-SOI CERBERUS micro-controller unit (MCU), depicted in Fig. \ref{Fig_4}(a). Detailed in Fig. \ref{Fig_4}(b), the accelerator features a highly-parallel datapath inspired from \cite{Kneip_2023} with 2$\times$32kB local memory (LMEM) and \textit{im2col} acceleration. IMAGINE's configurable datapath enables 128b end-to-end data transfers in a pipelined manner, providing precision- and size-dependent routing between the data LMEMs and the charge-domain CIM-SRAM macro, which contains the model's weights. The dataflow of the entire accelerator is further discussed in Section IV. Beforehand, let us first focus below on the CIM-SRAM architecture.

\subsection{Overall Macro Architecture}
\begin{figure}[!t]
	\centering
	\includegraphics[width=0.5\textwidth]{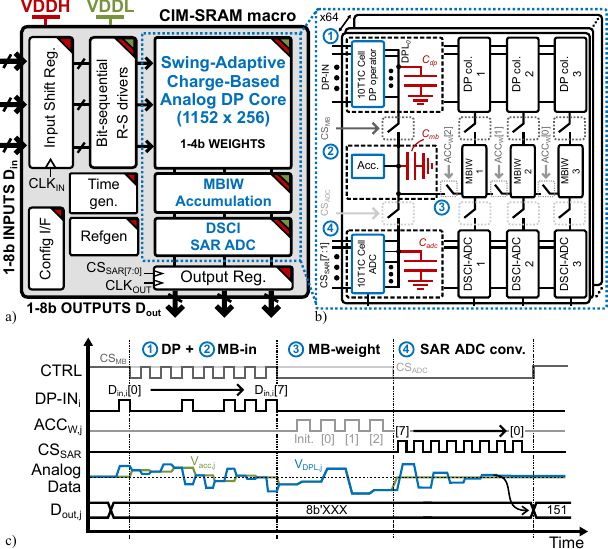}
	\caption{a) Top-level view of the charge-domain 1152$\times$256 CIM-SRAM macro, highlighting blocks of main interest. b) Coarse architecture of the 64 DP-to-ADC analog cores, mapping up to 4b weights and carrying out charge-based operations on the same sub-divided DPL along the way. c) Qualitative depiction of the macro's main operations.}
	\label{Fig_5}
	\vspace{-0.4cm}
\end{figure}

Fig. \ref{Fig_5}(a) showcases the top-level block diagram of the proposed CIM-SRAM macro, supporting 1-to-8b I/O CIM data and a conventional SRAM read/write (R/W) interface to store weights and biases (not shown here for convenience). The macro uses low and high voltage levels $V_{DDL}$ and $V_{DDH}$, respectively with nominal values of 0.4V and 0.8V. The analog core of the macro spans from the 1152$\times$256 DP array to the distribution-shaping charge-injection (DSCI) ADCs that implement the ABN function. Interestingly, both units involve the same 10T1C bitcell structure shown in Fig. \ref{Fig_2}(b). Between both ends, the multi-bit input-and-weight (MBIW) units carry out a bitwise sequential input accumulation before applying the weight bits by means of a summation across adjacent columns. Therefore, the analog core is split into 64 blocks of four columns each, mapping 1-to-4b weight bits as needed and plainly expendable to 8b. Importantly, Fig. \ref{Fig_2}(b) underlines that each colunm-wise analog operation occurs on the same DPL without discontinuity, from the DP computation down to the ADC conversion, relying on process-robust charge-domain operations along the whole way. Transmission gates progressively disconnect parts of the DPL as they become irrelevant, successively reducing the capacitive load seen during the DP, MBIW and ADC stages.


The macro's overall flow of operations can be divided into four main steps, qualitatively represented in Fig. \ref{Fig_5}(c), linked to the analog core structure. Assuming an 8b input precision $r_{in}$, unsigned input data are first broadcast along the enabled DP-IN(b) lines in parallel, performing charge-based DP operations as per Eq. (\ref{Eq_1}) one bit at a time. This process starts with LSB input bits and is repeated $r_{in}$ times, with each subsequent DP operation separated by an accumulation of the DP result on node $V_{acc}$ through a charge sharing with the DPL, as detailed in Section III.C, and a precharge phase of the DPL to $V_{DDL}$. Once all input bits have been processed, the DP array is entirely disconnected from the DPL and charge sharing between adjacent columns leads to the spatial accumulation of the weight bits, as foretold. Finally, ADC conversion takes place together with the ABN gain and offset stages, producing the final 1-to-8b rescaled output. The flexible computing precision of the macro allows a configurable trade-off between speed, energy and accuracy.

\subsection{Swing-Adaptive Charge-based DP Operator} 
\begin{figure}[!t]
	\centering
	\includegraphics[width=0.5\textwidth]{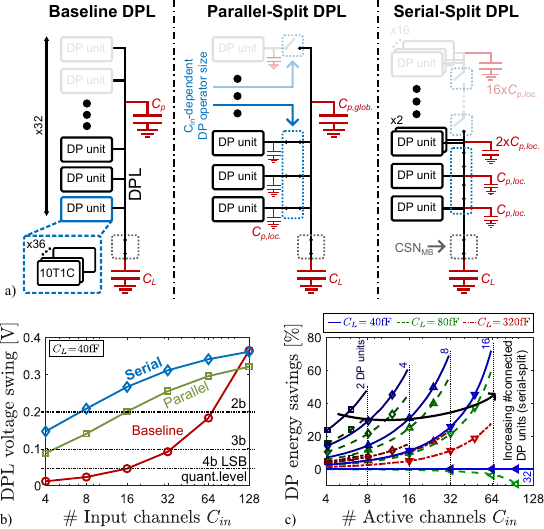}
	\caption{a) Comparison between baseline and split-DPL architectures, dividing the 1152 DP rows array into 32 even-sized units. b) Improvement of the DPL voltage swing with split-DPL techniques. c) DP energy savings for various capacitive loads in the serial-split DPL case.}
	\label{Fig_6}
\end{figure}

Within each column of the DP array, the total 1152 bitcell rows can be divided into 32 DP units containing 36 cells each. In this way, each unit can map a filter of 2D-convolutional layers with a kernel size of 3$\times$3 and a minimum input channel depth $\mathrm{C}_{in}$ of size 4. However, compared to the ideal operator in Eq. (\ref{Eq_1}), the presence of significant load capacitances on the DPL compresses the maximum DPL swing, replacing $\alpha$ by an actual attenuation factor $\alpha_{eff}$, given by

\begin{equation}
	\alpha_{eff} = C_c / (N_{dp} \; C_c + C_p + C_L),
	\label{Eq_3}
\end{equation}

\noindent where $N_{dp} = N_{rows}$ in baseline designs, $C_p$ models the parasitics due to metal routing, and $C_L$ is the total load capacitance related to non-DP blocks connected to the DPL. While $C_L$ is dominated by the ADC input capacitance, its total value can be brought down to 40fF by adapting the ADC architecture, as later described in Section III.D. To further improve $\alpha_{eff}$, one should thus try to make $N_{dp}$ and $C_p$ functions of the input channel depth $\mathrm{C}_{in}$ so as to match the CNN layer size. To that end, we present two ways to split the DPL in Fig. \ref{Fig_6}(a): a \textit{parallel-split} DPL with a $\mathrm{C}_{in}$-dependent amount of local DPLs connected to a shared global DPL through switches, and a \textit{serial-split} solution directly separating the sub-units with switches on the main DPL. Fig. \ref{Fig_6}(b) demonstrates the DPL swing improvement using both techniques compared to the baseline case, vastly restoring the effective number of ADC bits at low $\mathrm{C}_{in}$. Still, the maximum DPL swing remains limited by the significant DPL load $C_L$. This point is considered during CNN training to compensate the swing reduction by increasing the ABN gain during the DSCI-ADC stage. Furthermore, parallel-DPL suffers from additional parasitics $C_{p,glob}$ associated with the global DPL routing, which results in a lower swing improvement compared to a serial-split DPL. Finally, Fig. \ref{Fig_6}(c) highlights the DP energy reduction with such a DPL division as a function of the number of activated 3$\times$3 channel rows, for various $C_L$ loads and $\mathrm{C}_{in}$ configurations (i.e., number of connected DP units). The savings reach up to 72\% with 64 channels and a 40fF load, but rapidly diminish with a higher load as the total capacitance seen by the DP-IN driver rises. This further underlines the interest of optimizing the total DPL load.

\begin{figure}[!t]
	\centering
	\includegraphics[width=0.5\textwidth]{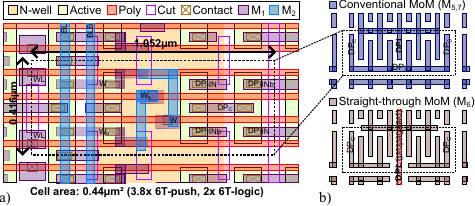}
	\caption{Layout of the implemented 10T1C bitcell, respectively showing a) the transistors position and b) the custom MoM atop the cell. The metal-6 (M6) MoM layer propagates the DPL vertically.}
	\label{Fig_7}
	\vspace{-0.4cm}
\end{figure}

The 10T1C bitcell's layout is drawn in Fig. \ref{Fig_7} and achieves an area of 0.44$\mu$m$^2$, corresponding to four times that of a 6T-pushed rule bitcell in the same technology. With bottom metal layers M1-3 used for the bitcell data, control and power routing, a custom MoM capacitor is layouted atop the cell in M5-6-7 layers, with the M6 one propagated vertically. The M4 layer remains mostly unused to avoid any significant coupling between the MoM and the internal bitcell nodes. The resulting $C_c$ capacitance reaches 0.7fF, generating a 2.4mV $k_B T/C_c$ noise from the active bitcell transistors. This noise remains below the 8b LSB voltage (3.125mV) and is further attenuated by $\alpha_{eff}$. Importantly, the metallization constraint does not allow to layout a parallel-split DPL without resorting on lower metal levels that would tightly couple with other signals. This point eventually motivates the use of the serial-split DPL solution.

\begin{figure}[!t]
	\centering
	\includegraphics[width=0.5\textwidth]{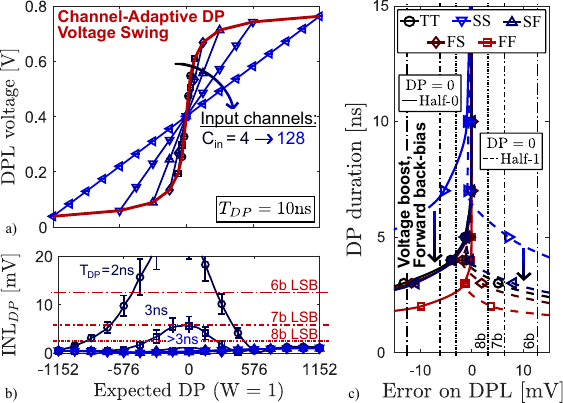}
	\caption{a) Post-layout transfer function of the DP operator, highlighting the configurable voltage swing. b) Linearity error INL$_{DP}$ observed for different DP durations $T_{DP}$. c) Worst-case error on the DP result for different process corners and with opposite worst-case weights distribution.}
	\label{Fig_8}
	\vspace{-0.4cm}
\end{figure}

The post-layout transfer function of the serial-split DP operator is reported in Fig. \ref{Fig_8}(a) with a 10ns DP duration $T_{DP}$, including device-to-device variability and highlighting the nonlinear dependence of the DPL swing on $\mathrm{C}_{in}$. Moreover, we consider in Fig. \ref{Fig_8}(b) the linearity error on the DP result $\mathrm{INL}_{DP} = | V_{DPL} - V_{lin} |$ as a function of the DP duration across the full array size. Here, $V_{lin}$ is given by Eq. (\ref{Eq_1}), with $\alpha = \alpha_{eff}$. Indeed, the transmission gates in the serial-split DPL architecture limit the charge-sharing speed, especially as the target DPL voltage nears $V_{DDH}/2$, weakening the driving strength of these gates. This increases the DP duration needed to satisfy the worst-case DPL settling time, or leads to a case-dependent error on the DPL voltage. The largest errors are obtained with the bottom half of the DP array injecting a positive voltage (\textit{half-1}) onto the DPL, while the top half injects a negative one, or the opposite (\textit{half-0}). This phenomenon is worsened when accounting for process corners as in Fig. \ref{Fig_8}(c), requiring margins on the DP duration. Here, we choose a duration of 5ns per single-bit DP, with a +/-1ns configurability range. In that way, we achieve an acceptable speed while limiting the linearity error below one LSB, and model it during CNN training. Note that the parallel-split DPL only needs a 1.5ns delay thanks to its lower series resistance.



\begin{figure}[!t]
	\centering
	\includegraphics[width=0.5\textwidth]{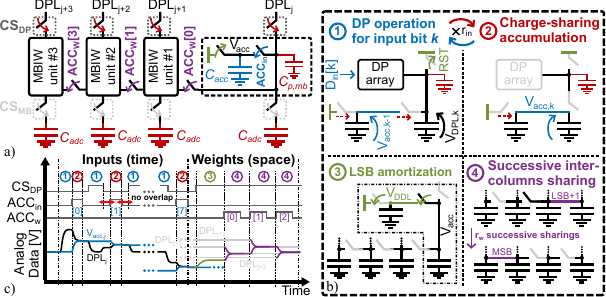}
	\caption{a) Architecture of the MBIW unit, based on iterative charge-sharing both in time (input) and space (weights). b) Sequence of operations of the MBIW unit, with c) qualitative depiction of the sequential-input, spatial-weight accumulation.}
	\label{Fig_10}
	\vspace{-0.4cm}
\end{figure}

\subsection{Multi-Bit Input-and-Weight Accumulation}

The overall architecture of one of the 64 MBIW units is depicted in Fig. \ref{Fig_10}(a) across a block of four adjacent columns. Each column possesses its own accumulation capacitance $C_{acc}$, layouted within the vertical 10T1C pitch and sized to equal the remaining DPL load. Hence, $C_{acc} = C_{mb} + C_{adc}$, with $C_{mb}$ and $C_{adc}$ respectively the total DPL loads associated with the MBIW and ADC blocks. The MBIW operates along four phases described in Figs. \ref{Fig_10}(b) and (c), all relying on capacitive charge-sharing. Phases 1 and 2 correspond to the accumulation of input bit precisions by repeating the DP operation $r_{in}$ times. During the DP phase, the DPL voltage is sampled on the total MBIW and ADC load capacitance, while the accumulation capacitance is disconnected from it, storing the previous accumulation voltage $V_{acc,k-1}$. Then, the DP array is disconnected, while the accumulation capacitance is shared with the DPL load. Before performing the next DP computation, the MBIW accumulation node is disconnected from that load and the DPL is reset to $V_{DDL}$. Critically, the signals that respectively control connections to the DP array (CS$_{DP}$) and accumulation node (ACC$_{in}$) must not overlap to avoid any corruption of the stored $V_{acc}$. The DPL voltage resulting from these $r_{in}$ cycles is given by


\vspace{-0.1cm}

\begin{equation}
	\begin{split}
	V_{DPL} = V_{DDL} \, \bigg( \alpha_{mb}^{r_{in}-1} + & \\
	 \suml_{k=0}^{r_{in}-1} \alpha_{mb}^{r_{in}-k}  \; \big(1 \; + & \;  \alpha_{eff} \suml_{i=0}^{N_{on}-1} \mathrm{X}_i[k] \, \mathrm{W}_{i,j} \big) \bigg)
	\end{split}
	\label{Eq_4}
\end{equation}

\noindent where $\alpha_{mb} = (C_{mb}+C_{adc})/(C_{acc}+C_{mb}+C_{adc}) \simeq 1/2$ is the multi-bit attenuation factor. In case of binary inputs, the accumulation phase is bypassed altogether, preserving the voltage swing seen at DP time. Once the input accumulation is finished, weight accumulation takes place during phases 3 and 4. The LSB weight is first self-weighted by sharing its DPL with the $V_{DDL}$-precharged accumulation node. Then, inter-column charge sharing takes place, successively sharing DPLs by pairs of two from LSB to MSB, obtaining the final MBIW result on the MSB's DPL

\vspace{-0.1cm}

\begin{equation}
	V_{MBIW} = \suml_{k=0}^{r_w-1} (1/2)^{r_w-k} \; V_{DPL,k}
	\label{Eq_5}
\end{equation}

\noindent with $r_w$ the weights precision. Observe that performing inter-column charge-sharing by pairs of two rather than all at once mitigates the voltage range compression  resulting from the sharing. Also, extending this scheme to 8b weights is straightforward by connecting more columns in a similar manner. 8b support is missing here to limit the configuration complexity.

\begin{figure}[!t]
	\centering
	\includegraphics[width=0.5\textwidth]{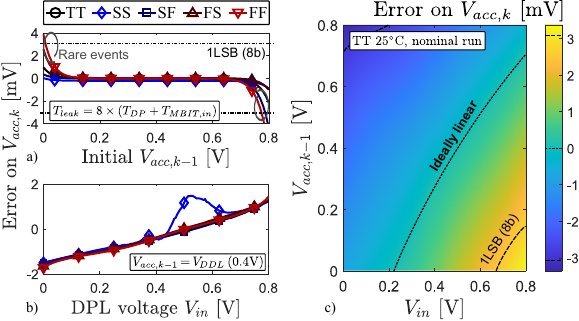}
	\caption{Linearity error on the accumulation voltage across process corners due to a) leakage currents as a function of the initial $V_{acc,k-1}$, b) charge-injection, for different MBIW input voltages $V_{in}$. c) 2-D error map on the on the input accumulation voltage $V_{acc,k}$, depending on both the input voltage from the $k$-th DP operation and the previously stored voltage $V_{acc,k-1}$.}
	\label{Fig_11}
	\vspace{-0.4cm}
\end{figure}

\begin{figure*}[!t]
	\centering
	\includegraphics[width=\textwidth]{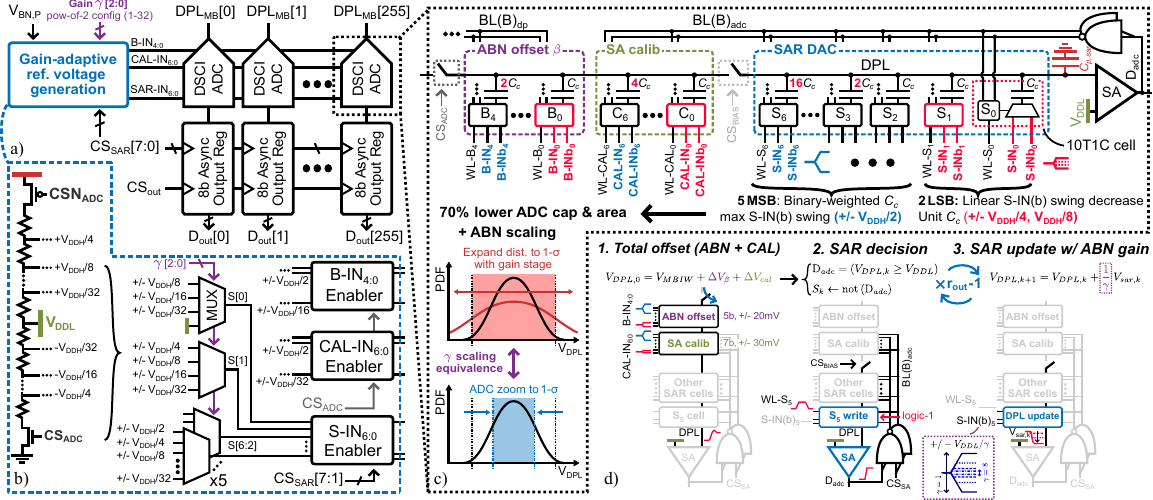}
	\caption{a) Overall architecture of the distribution-shaping charge-injection (DSCI) ADCs, and of b) its shared gain-adaptive reference ladder implementing ADC thresholds zooming. c) Details of the 10T1C-based DSCI ADC architecture, with d) qualitative depiction of its operations during data conversion. From the data's point of view, ADC zooming taking place during the conversion is equivalent to a scaling effect.}
	\label{Fig_12}
	\vspace{-0.4cm}
\end{figure*}

Aside from the below-1\% capacitance imbalance described by $\alpha_{mb}$, the high linearity of Eqs. (\ref{Eq_4}) and (\ref{Eq_5}) is altered by leakage currents that (dis)charge the accumulation capacitance, and by non-zero charge injections from the MOS transmission gates serving as switches. Figs. \ref{Fig_11}(a) and (b) respectively consider the impact of both non-idealities on node $V_{acc}$ during input accumulation, across all process corners. Regarding leakage, the voltage error measured at the end of the 8b input accumulation phase $T_{leak}$ remains negligible, except for unlikely extreme voltage values. Besides, while the impact of charge injection changes with the MBIW input voltage $V_{in,k}$ (corresponding to the $k$-th DP result), it stays below one LSB across all process corners. This dependence on voltage comes from the change in gate-to-source/drain capacitances of the transmission gates with their $V_{gs}$ voltage, such that the error on $V_{acc,k}$ depends on both the input value $V_{in,k}$ and the accumulation voltage $V_{acc,k-1}$ stored before the charge-sharing step. Fig. \ref{Fig_11}(c) showcases a 2D-map of this error in nominal conditions, highlighting the zero-error curve. The accumulation error reaches up to +/-1 LSB of an 8b ADC, acting as a deterministic input-dependent offset that we model during the CNN training.


\subsection{Distribution-Shaping Charge-Injection ADC}
At the end of the bitwise accumulation, the MBIW unit is disconnected from the DPL and column-wise ADCs with a 1-to-8b precision $r_{out}$ transform the analog DP result stored on the DPL into digital outputs $\mathrm{D}_{out}$. As seen in Fig. \ref{Fig_12}(a), these outputs are stored within custom 8b asynchronous DFF registers. Relying on separated control signals for their master and slave latches allows to simultaneously write new ADC outputs to the first latch while preserving previous values in the second latch. Consequently, data available at the CIM output are only updated at the next positive clock edge by a $\mathrm{CS}_{out}$ pulse, which enables system-level pipelining as later discussed in Section IV. Finally, the necessary ADC inputs are generated by a gain-adaptive unit based on process- and variation-insensitive reference voltages $V_{BN,P}$, generated by an on-chip calibrated reference \cite{Lefebvre_tcas_2022}. 

Detailed in Fig. \ref{Fig_12}(c), the DSCI ADCs perform data conversion in the charge domain, similar to the rest of the analog core. On top of providing in-situ storage ability, 10T1C bitcells form a binary-weighted capacitive array that mimics a charge-injection DAC, where the DPL holds the residual voltage in a SAR-like conversion. Moreover, the ADC performs distribution shaping by implementing ABN offset and gain stages. Hence, the overall ADC is divided in three sub-blocks: (i) a 5b ABN offset unit achieving a +/-30mV offset range on the DPL, (ii) a 7b calibration unit to counteract the sense amplifier (SA) offset, and (iii) an 8b SAR array embedding the ABN gain function. As such, the ADC's 10T1C bitcells not only hold the pre-computed offset and calibration data, but also duplicate the SAR output code storage in-situ. This redundancy avoids to face extreme routing congestion between the 8b ADC registers and the SAR capacitive bank within the tight column pitch.

\begin{figure}[!t]
	\centering
	\includegraphics[width=0.5\textwidth]{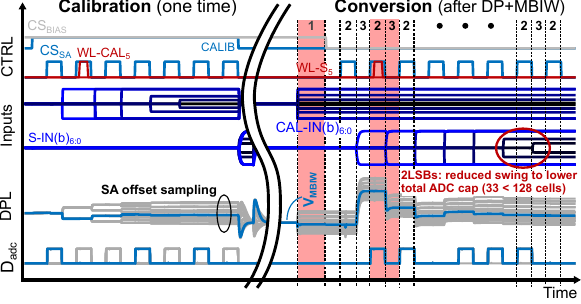}
	\caption{Post-layout operations of the calibration and conversion phases of the ADC ($\gamma = 1$), showcasing 100 Monte-Carlo iterations.}
	\label{Fig_13}
	\vspace{-0.4cm}
\end{figure}

On top of exploiting bitcells, the ADC introduces a voltage-split charge-injection DAC topology, seen in \ref{Fig_12}(c), that concurrently reduces the total ADC load and enables ABN scaling. Here, relying on a conventional capacitance-split DAC \cite{Lee_dac_2021} to shrink the large 8b load $C_{adc} = 128 \times C_c$ is prohibited by the floating state of the DPL, which is sensitive to kickback coupling. Instead, the SAR array is split into an MSB part with binary-weighted capacitance values, similar to a conventional DAC, and an LSB part with unit capacitance but linearly-downscaled input swing on the S-IN(b) lines. In this design, the LSB array uses two bitcells to reduce the ADC load by more than 70\%, while only requiring two additional input levels and bringing a negligible amount of additional nonlinearity. This technique is also used in the offset and calibration blocks, improving their maximum swing on the DPL by minimizing the load overhead. As a result, ADCs account for less than 5\% of the total CIM-SRAM area, with a total load of 40fF per column, including parasitics, enabling the high energy savings predicted in Fig. \ref{Fig_6}(c). Furthermore, changing the maximum voltage swing of the SAR inputs also offers the opportunity to perform global ABN scaling without the need for an explicit gain stage. Indeed, applying the invert gain factor $1/\gamma$ to all S-IN(b) lines compresses the dynamic range of the ADC, working as a zoom effect equivalent to an explicit scaling of the DPL voltage distribution, as depicted in Fig. \ref{Fig_12}(c). To that end, the reference unit detailed in Fig. \ref{Fig_12}(b) adapts the S-IN(b) levels feeding the MSB and LSB split DAC arrays based on the $\gamma$ configuration. These levels are directly selected from a double-sided resistive ladder, activated during the ADC operation and drawing a 1mA current to settle all S-IN(b) lines within 5ns. Due to mismatch and area constraints, the ladder affords a minimum voltage step of $V_{DDH}/32$, such that the MSB split DAC achieves a maximum gain of 16.

The post-layout operations of the resulting ADC are reported in Fig. \ref{Fig_13}, both in calibration and conversion modes. Focusing on the conversion mode first, happening at the end of each CIM cycle, ADC operations occur in three phases, depicted in Fig. \ref{Fig_12}(d). First, the ABN offset and SA calibration blocks add an offset to the initial DPL voltage, following their pre-stored bitcells data. These bitcells are then disconnected from the residual DPL, and SAR operations start. The \mbox{$k$-th} conversion cycle of the SAR begins with a binary SA decision on the DPL voltage to obtain the corresponding \mbox{$k$-th} output bit, storing it inside both its column register and the corresponding split-DAC bitcell. Next, the SAR residue is computed by updating the DPL based on the stored value, selecting either S-IN$_k$ or S-INb$_k$ to inject charges on the DPL. As described earlier, the ABN zoom effect is applied during this stage by downscaling the S-IN(b) voltage swing. The decision and update phases are repeated $r_{out}$ times, increasing the SAR delay linearly and, to some extent, its conversion energy. Eventually, the resulting ADC output is given by

\begin{figure}[!t]
	\centering
	\includegraphics[width=0.5\textwidth]{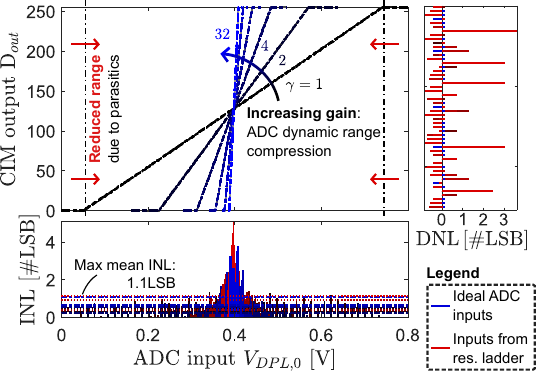}
	\caption{Nominal post-layout transfer function of the ADC for increasing values of gain $\gamma$. The nominal DNL and INL of the ADC increase with $\gamma$ due to exacerbated sensitivity to voltage errors.}
	\label{Fig_14}
	\vspace{-0.4cm}
\end{figure}

\vspace{-0.1cm}

\begin{equation}
	\mathrm{D}_{out} = \rho\left(2^{r_{out}-1} + \gamma \, \dfrac{\Delta{V_{MBIW}} + \Delta{V_\beta} + \Delta{V_{cal}}}{\alpha_{adc} \, \left(V_{DDH}/2^{\,r_{out}-1}\right)}\right),
	\label{Eq_adc}
\end{equation}

\noindent where $\rho(\,)$ is the integer floor function, $\Delta{V_\beta}$ and $\Delta{V_{cal}}$ are the ABN bias and calibration offsets, and $\alpha_{adc} = C_{sar} / (C_{sar} + C_{p,sar})$ accounts for the total capacitance of the SAR array, with $C_{sar} = 33 \, C_c$. The nominal ADC transfer function extracted in Fig. \ref{Fig_14} with zero offset and calibration confirms the high linearity of the ADC and demonstrates the impact of ABN gain scaling. Firstly, we observe that the input dynamic range is compressed due to the attenuation effect, which suitably brings the ADC range closer to the maximum DP voltage range, previously compressed during the DP and MBIW steps. More critically, both the nominal INL $(= | \mathrm{D}_{out} - \mathrm{D}_{lin} |)$ and DNL $(=  \mathrm{D}_{out,k} - \mathrm{D}_{out,k-1} \forall \; k \in [1,2^{r_{out}}[)$ of the ADC increase with a higher gain $\gamma$, as the sensitivity to small nonlinearities between subsequent charge injections increases. These nonlinearities are worsened when considering the parasitic resistances and the mismatch of the resistive ladder, distorting the perfect linearity of the voltage steps and thereby leading to lost LSB information above $\gamma = 8$, as discussed before. In the end, the ADC reaches a mean INL error of 1.1 LSB, while its peak error rises up to 4.5 LSB when $\gamma = 32$, as the LSB voltage step shrinks. While these limited errors can be deterministically known at CNN train time, both the DNL and INL are further worsened by stochastic sources, namely the SA mismatch as well as the intrinsic noise affecting all DPL-connected blocks. Therefore, we purposefully added a calibration mechanism to the ADC.   

\subsection{Mismatch and Low-Frequency Noise Calibration}

\begin{figure}[!t]
	\centering
	\includegraphics[width=0.5\textwidth]{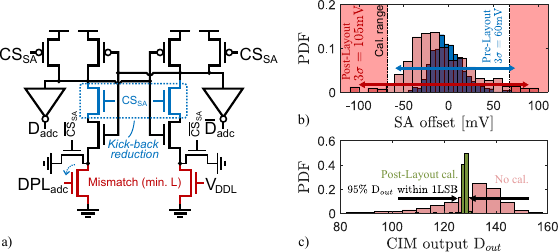}
	\caption{a) Architecture of the StrongArm SA, with a kickback reduction structure and minimum transistor lengths. b) Distribution of the SA offset, worsened post-layout do to layout resizing constraints and proximity effects. c) Calibration brings 95\% of the CIM outputs back within one LSB ($V_{DPL,0} = 0.4$V).}
	\label{Fig_15}
	\vspace{-0.4cm}
\end{figure}

The ADC calibration occurs on a rare basis to refresh the combined compensation of the SA mismatch and the low-frequency noise affecting the DPL. During calibration, $V_{DPL,0}$ is precharged to $V_{DDL}$ while decision and update phases happen similarly to the conversion phase, but applied to the calibration unit instead of the SAR. In this way, the calibration code converges towards matching the voltage offset seen at the input of each SA, compensating for it during CIM computations as seen in Fig. \ref{Fig_14}. This offset is namely dominated by transistor mismatch within the SA, which implements the low-kickback StrongArm topology with minimum-length input differential pair shown in Fig. \ref{Fig_15}(a). These techniques minimize the kickback level on the floating DPL undergone during each SA decision, bringing it below 0.03mV. However, the minimum-length devices further degrade the robustness to mismatch. Therefore, the 7b calibration unit adopts a $4 \times C_c$ MSB capacitance, covering the 3-$\sigma$ 60mV pre-layout SA offset observed in Fig. \ref{Fig_15}(b). As such, the calibration process offers a resolution of 0.47mV, making the SAR resilient to any low-frequency SA offset for $\gamma < 8$. Yet, post-layout effects and resizing constraints to fit the column pitch lead to a steep 75\% increase of the pre-layout deviation, such that only a 2-$\sigma$ offset range remains fully handled by the calibration block. With no design-time compensation, out-of-bond offsets might lead to a few dysfunctional columns per macro, as seen in Fig. \ref{Fig_15}(c). If identified, extreme offsets can be partly dealt with by the ABN offset block, reducing the tunable offset range for that column. 

\section{CIM-CNN Accelerator Dataflow}
\externaldocument{cimu_architecture}

\begin{figure*}[!t]
	\centering
	\includegraphics[width=\textwidth]{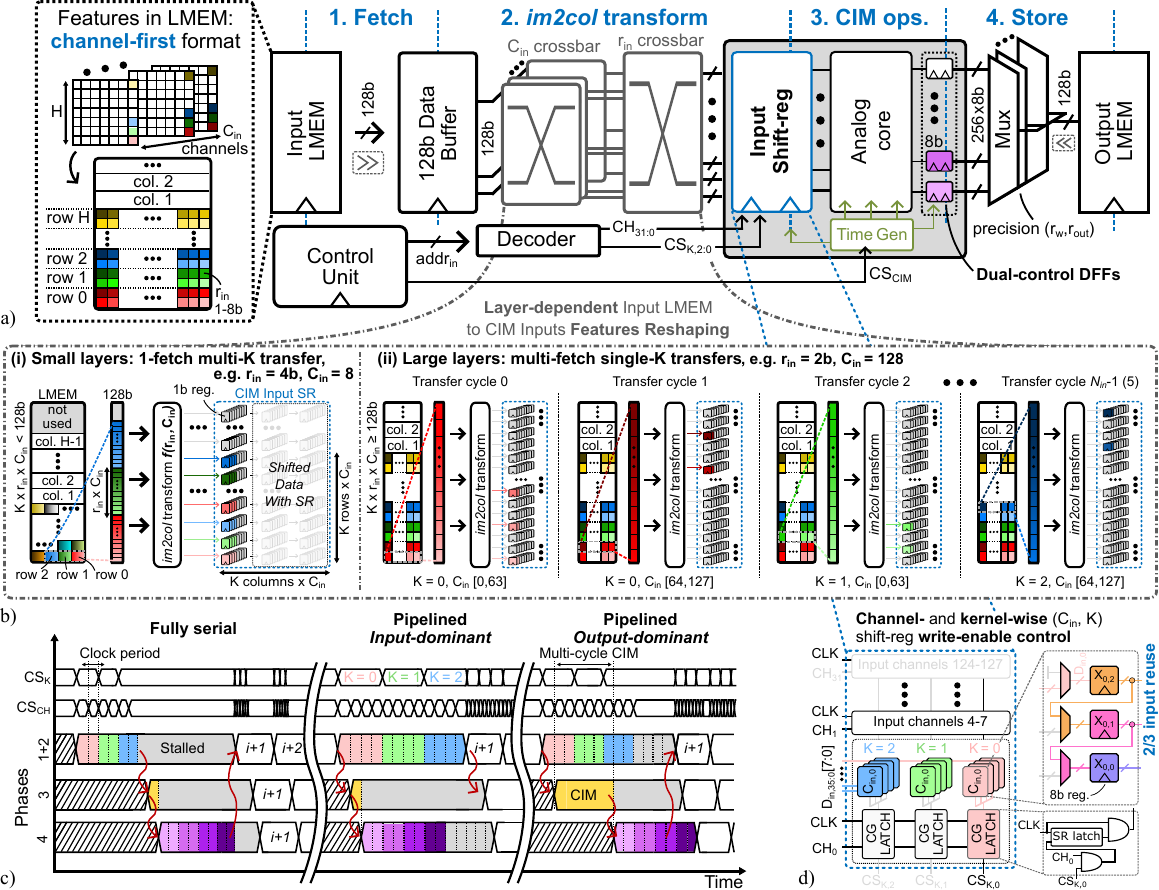}
	\caption{a) Detailed architecture of the IMAGINE accelerator, with channel-first input LMEM data and a 128b four-stage pipelined datapath. (b) Illustration of data transfers and \textit{im2col} reshaping between the input LMEM and CIM-SRAM input registers for (i) few-channel and (ii) many-channel convolutional layers, respectively fetching the data of multiple or a single kernel(s) per transfer. (c) Qualitative comparison of the accelerator's operations in serial and pipelined cases, with both input- and output-dominated LMEM transfers. (d) Detail of the conditionally-updated CIM-SRAM's input shift-register, which enables more than 60\% digital area reduction by avoiding one-shot im2col and data pre-buffering.}
	\label{Fig_16}
	\vspace{-0.4cm}
\end{figure*}

IMAGINE embeds the CIM-SRAM macro within a 1-to-8b highly parallel datapath to provide layer-by-layer execution of CNNs. The digital datapath surrounding the macro is based on \cite{Kneip_2023} and represented in Fig. \ref{Fig_16}(a): it operates with 128b I/O transfers between the CIM-SRAM and LMEMs regardless of the configured bit precision and number of channels. The accelerator operates along four pipelined stages: (i) data fetching from the input LMEM, where data are encoded in a precision-first, channel-second and kernel-last format (ii) optional \textit{im2col} transform for convolutional layers, rearranging the input data in a channel-last format suiting the CIM-SRAM's input shift-register, and applying zero-padding as requested, (iii) CIM computation as described in Section III, and finally (iv) CIM output storage to the output LMEM, in the same kernel-last format. Such format not only minimizes the number of LMEM accesses, but also enables direct data reuse for the next layer, after switching input and output LMEMs in a ping-pong manner. In that sense, LMEM data mapping is similar to \cite{Kneip_2023}, extending here the support to an 8b I/O precision. Besides, stage (ii) supports an optional signed-to-unsigned type conversion, and stage (iv) its reverse. While being optimized for 3$\times$3 kernels, larger ones (5$\times$5, etc.) can be executed on IMAGINE as a succession of 3$\times$3 kernels \cite{Simonyan_2014}. 

Compared to \cite{Kneip_2023}, the $im2col$ conversion is performed sequentially on batches of 128b data fetch from the input LMEM, rather than in a one-shot fashion. This change significantly reduces the required size of the pre-$im2col$ buffer, down to 128b instead of the CIM-SRAM's $1152 \times 8$b full input bandwidth, reducing the bit-normalized area overhead of the digital datapath by more than 60\%. Nonetheless, this solution requires a more complex input shift-register architecture to support the conditional enabling of different input register subsets on the CIM-SRAM side, depending on the target layer configuration. To that end, the entire shift-register is split into 32 sub-block as detailed in Fig. \ref{Fig_16}(d), matching the DP array division described in Section III.A. Within each sub-block, local clock gating (CG) latches control the update of the block's registers, with 32 CH$_i$ signals dictating the access to each block, while three CS$_{K,j}$ signals decide which kernels within the blocks are to be updated. Therefore, the minimum configuration of the CIM-SRAM macro is four input channels in convolutional mode, and one full sub-block in fully-connected mode.

To illustrate the digital-to-macro interface functionality, two transfer situations between the input LMEM and the CIM-SRAM's shift-register are covered in Fig. \ref{Fig_16}(b), highlighting the \textit{im2col} data reshaping in convolutional mode. On the one hand, CNN layers with few channels and low precision can transfer multiple kernels in a same image row within a single transfer. On the other hand, large CNN layers need to split the transfer of a single kernel over multiple cycles. At fixed $\mathrm{C}_{in}$, the number of cycles is directly proportional to the input precision $r_{in}$, re-routing the shift-register inputs in the \textit{im2col} as depicted.

	 
Eventually, Fig. \ref{Fig_16}(d) presents the qualitative flow of IMAGINE's operation phases, considering various situations. Assuming a serial processing first, input transfers remain stalled until all CIM output data have been stored to the output LMEM, leading to a cycle penalty of

\vspace{-0.1cm}

\begin{equation}
	N_{stall} = 1 + N_{cim} + \mathrm{ceil} \left( \dfrac{r_{out} \times C_{out}}{\mathrm{BW}} \right),
	\label{Eq_stall}
\end{equation}

\noindent where $\mathrm{BW} = 128$b is the LMEM I/O bandwidth and $N_{cim}$ is the number of clock cycles allowed for CIM-SRAM operations, usually set to one. This penalty severely hinders the overall CIM-CNN throughput, calling for pipelining IMAGINE's operating phases. Ideally, all four phases might then take place simultaneously: while the CIM-SRAM performs the $i$-th DP computation, previous outputs $i-1$ are stored to the output LMEM whereas new inputs $i+1$ are fetched from the input LMEM. However, this pipeline is subject to data-movement constraints between the different phases, depending on the CNN layer configuration, similar to \cite{Kneip_2023}. On the one hand, \textit{input-dominated} layers wait for input transfers to complete before issuing the next CIM-SRAM and store operations. For a convolutional layer, the number of cycles needed to process a single output-map value is then given by

\vspace{-0.1cm}

\begin{equation}
	N_{cycles} = N_{in} = (N_{cim}-1) + \mathrm{ceil} \left( \dfrac{K \times r_{in} \times C_{in}}{\mathrm{BW}} \right).
	\label{Eq_N_in}
\end{equation}

\noindent Eq. (\ref{Eq_N_in}) showcases that multi-cycle CIM-SRAM computations increase the number of cycles as the data stored within the input shift-register have to remain constant during the entire macro operation. Relying on split control signals for the master and slave latches, as for the macro's output registers, could however circumvent this bottleneck. Moreover, Eq. (\ref{Eq_N_in}) only holds for data transfers within a same image row, dividing the number of transfers per $K$ thanks to the input shift register. At the start of a new row, $K \times N_{in}$ cycles are necessary to fetch the whole new input data kernel. On the other hand, \textit{output-dominated} layers stall new input data fetches and CIM-SRAM operations while outputs remain to be transferred. In that case, the number of cycles is given by 

\vspace{-0.1cm}

\begin{equation}
	N_{cycles} = N_{out} = N_{cim} + \mathrm{ceil} \left( \dfrac{r_{out} \times C_{out}}{\mathrm{BW}} \right) - 1.
	\label{Eq_N_out}	
\end{equation}

\noindent Now, multi-cycle CIM-SRAM operations delay the start of new output storages, which stresses the importance of minimizing the CIM-SRAM's operation time so as to avoid hindering the overall accelerator performance.

Finally, one may wonder how the map workloads that do not fit within the available LMEM or CIM-SRAM capacity. In such cases, the overall execution of a single CNN layer is split into CIM-CNN processing phases, as described above, and CIM-CNN read/write phases to move weights (resp. intermediate results) between an off-chip DRAM and the on-chip CIM-SRAM (resp. LMEMs). These read/write phases pose an end-to-end latency and energy bottleneck on the overall CNN execution. The additional latency depends on the ratio between the off-chip bandwidth and the CIM-CNN bandwidth. For instance, with a 32b off-chip BW, the latency for weight transfers is nearly equivalent to the number of cycles needed to process a single image, assuming a same clock frequency. Regarding energy, 32b DRAM accesses for weight fetching would amount to a total overhead below 10\% over the whole image processing, which is acceptable. This overhead, however, may increase if the size of the intermediate layer maps scales down compared to that of the input image, due to pooling layers or striding. In such cases, providing more on-chip storage capacity is the go-to solution, at the expense of more area.

\externaldocument{cim_cnn_accelerator}
\section{Measurement Results}
\begin{figure}[!t]
	\centering
	\includegraphics[width=0.5\textwidth]{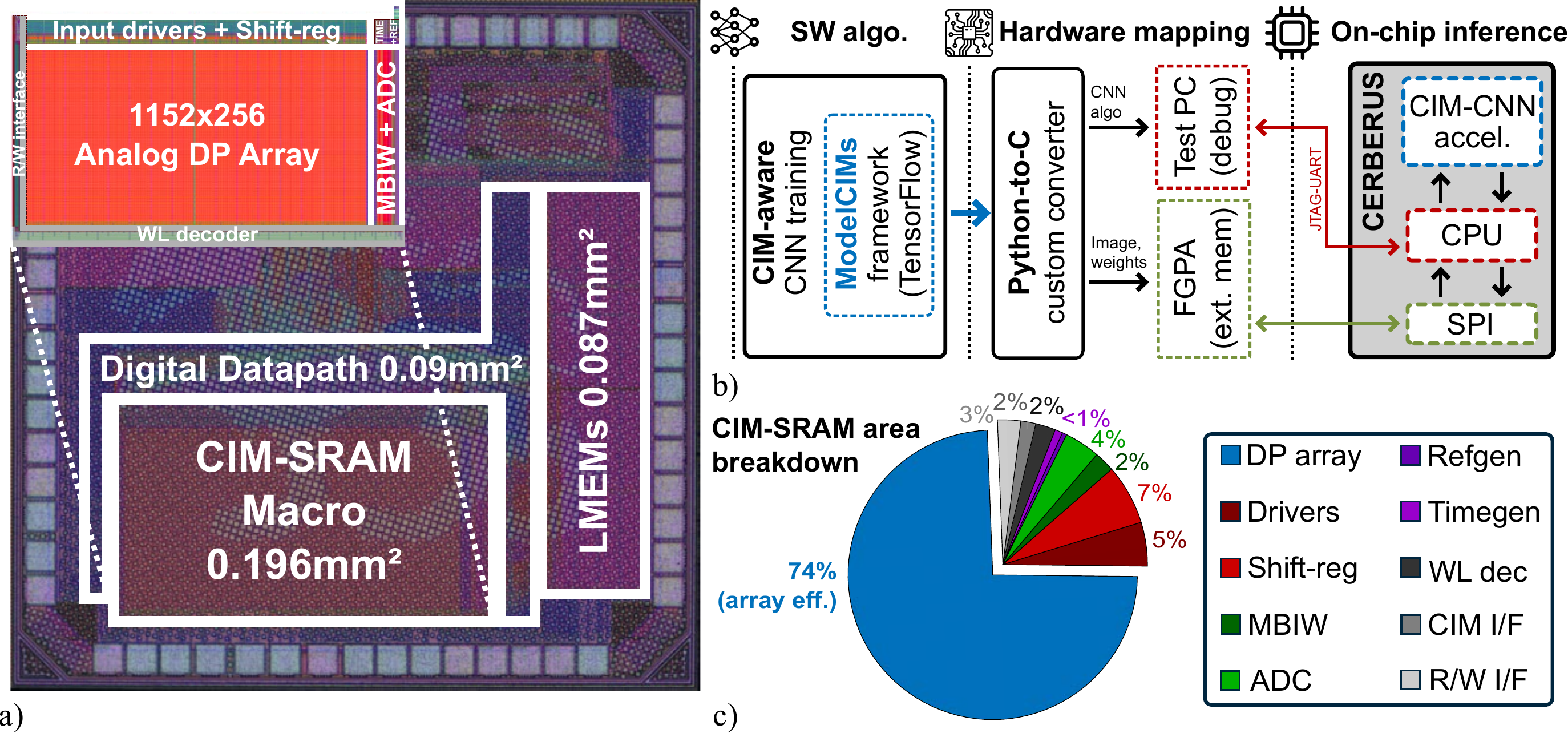}
	\caption{a) Chip microphotograph with CIM-SRAM macro layout. b) Measurement setup for CNN use cases. c) CIM-SRAM area breakdown.}
	\label{Fig_17}
	\vspace{-0.4cm}
\end{figure}

The IMAGINE accelerator was embedded in the CERBERUS MCU, fabricated in 22nm FD-SOI. The chip microphotograph and the macro's layout are shown in Fig. \ref{Fig_17}(a). The CIM-SRAM area is dominated at 74\% by the DP array, due to the use of large 10T1C bitcells while minimizing the ADC area as described in Section III.D. Overall, the macro occupies 53\% of the 0.373mm$^2$ total accelerator's area. Both the individual macro and entire accelerator have been characterized using the setup shown in Fig. \ref{Fig_17}(b), allowing to decouple macro- and system-level performance.

\subsection{CIM-SRAM Characterization}
\begin{figure}[!t]
	\centering
	\includegraphics[width=0.5\textwidth]{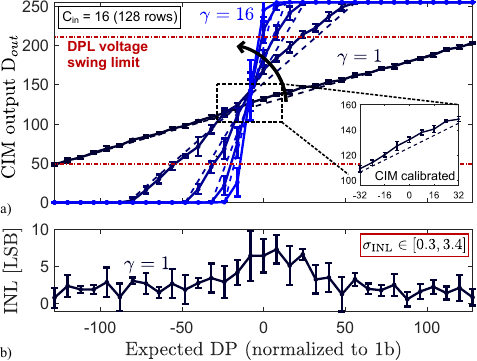}
	\caption{a) CIM-SRAM 8b tranfer function at 0.6V, with 16 input channels in FC mode and increasing gain $\gamma$. b) Observed INL at unity gain.}
	\label{Fig_18}
	\vspace{-0.4cm}
\end{figure}

\begin{figure}[!t]
	\centering
	\includegraphics[width=0.5\textwidth]{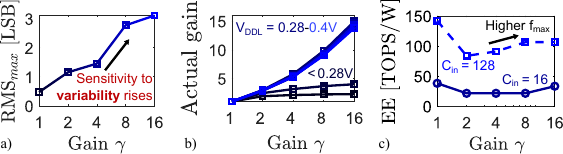}
	\caption{Impact of gain $\gamma$ scaling on a) the maximum output RMS error on the INL, b) the linearity of the gain for different supplies, c) the macro's 8b peak energy efficiency (EE).}
	\label{Fig_19}
	\vspace{-0.4cm}
\end{figure}

\begin{figure}[!t]
	\centering
	\includegraphics[width=0.5\textwidth]{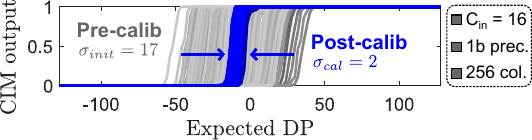}
	\caption{CIM-SRAM 1b input-referred deviation prior to and after SA offset calibration across 256 columns (average over 100 different samples).}
	\label{Fig_20}
	\vspace{-0.4cm}
\end{figure}

Firstly, standalone characterization of the CIM-SRAM macro is carried out using its fully-connected (FC) configuration to enable easy one-to-one mapping between input LMEM and CIM data. A dedicated test mode allows accurate power estimates by ensuring a 100\% duty-cycled utilization of the macro, changing a single 128b patch of inputs per CIM cycle. Fig. \ref{Fig_18} reports the macro's transfer function and INL at 8b and 0.6V measured across 256 columns, considering 16 channels (i.e., 128 rows in FC mode) and varying its gain $\gamma$. Here, inputs are kept at zero while weights are progressively changed from all-0's to all-1's, from the bottom to the top of the DP array. In this way, we can detect a peak of mean INL around zero-valued DPs, resulting from a slightly short DP timing pulse in the measured slow chip corner, as predicted in Fig. \ref{Fig_8}(b). Moreover, the aggregated output variability resulting from temporal noise (100 iterations) and residual column-to-column mismatch, after calibration, amounts to a maximum deviation of 3.5 LSB on the INL under unity gain. This yields a maximum RMS error of 0.52 LSB, which scales up with $\gamma$ in Fig. \ref{Fig_19}(a) given the growing sensitivity to the residual noise floor. This error level is obtained after performing the calibration described in Section III.D, which brings the spatial deviation from 17 LSB down to just 2 LSB at the 8b precision, as seen in Fig. \ref{Fig_20}. Residual errors come from of mix of out-of-range SA offsets, insufficient voltage resolution during the calibration process, and noise induced by MoM caps. Reducing the error further down requires to use larger $C_c$ capacitances and to upsize the SA to contain its mismatch. Nonetheless, both techniques would negatively impact the power and area of the macro. Instead, these low-noise statistics are included during the off-line CNN training to be partly compensated. Furthermore, with regards to $\gamma$ scaling, linearity slowly degrades as the $V_{DDL}$ supply voltage is reduced from 0.4 down to 0.28V in Fig. \ref{Fig_19}(b), keeping $V_{DDL} = V_{DDH}/2$. Functionality is lost below 0.28V due to the insufficient configuration range of internal timings. Finally, the 8b peak energy efficiency of the CIM-SRAM macro indirectly depends on $\gamma$ in Fig. \ref{Fig_19}(c), as the maximum operating frequency of the macro slightly improves between 2 and 16 thanks to the shorter transition time of the compressed $V_{sar}$ levels. However, the efficiency remains better for unity gain as the SAR MSBs are directly connected to ground and supply levels, alleviating the resistive ladder's total load.

\begin{figure}[!t]
	\centering
	\includegraphics[width=0.5\textwidth]{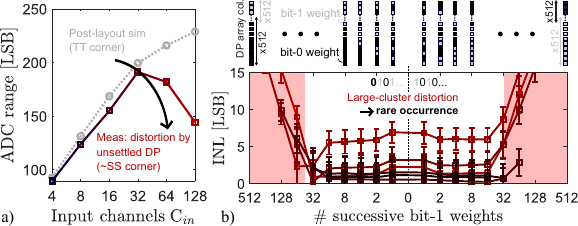}
	\caption{a) Measured increase of the mean ADC output range with $\mathrm{C}_{in}$ ($\gamma = 1$). Unsettled DP in the slow chip corner results in distortion at high $\mathrm{C}_{in}$. b) Measured distortion per $\mathrm{C}_{in}$ configuration for a zero-valued expected DP under incremental weight-value clustering (inputs fixed at zero). Mean INL strongly rises in rare highly-clustered cases.}
	\label{Fig_21}
	\vspace{-0.4cm}
\end{figure}

\begin{figure}[!t]
	\centering
	\includegraphics[width=0.45\textwidth]{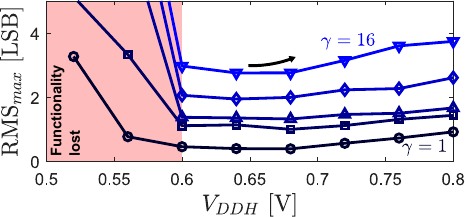}
	\caption{Under unity gain $\gamma$, the 8b output RMS error increases with supply voltage following timing pulses shortening and higher IR drops ($\mathrm{C}_{in} = 16$).}
	\label{Fig_22}
	\vspace{-0.4cm}
\end{figure}

Fig. \ref{Fig_21}(a) showcases the impact of changing the number of input channels used. Overall, the output dynamic range improves at constant gain and supply when scaling $\mathrm{C}_{in}$ up, closely following post-layout estimates up to 32 channels. However, the maximum swing drops above this value as a result of distortions resulting from an unsettled DP result in the measured slow chip corner. Such distortion is estimated in Fig. \ref{Fig_21}(b) by measuring the output obtained when expecting a zero-valued DP, realized through different combinations of weight and input values. In particular, with inputs fixed at zero and half of the weights storing a bit-1, the other half a bit-0, we increment the number of row-wise consecutive bit-1 (resp. bit-0) weights, starting with a bit-1 (resp. bit-0) in the first CIM-SRAM row. A significant increase in INL appears above 32 consecutive values due to opposing charge injections in the different sub-parts of the split-DPL array, which lacks enough time to properly settle the ADC input node in a slow chip corner. This issue would be avoided with a larger DP timing configuration or a parallel-split DP structure, which could both not be implemented here due to tight area constraints and metallization limitations. Eventually, Fig. \ref{Fig_22} points out that the maximum RMS error slightly increases with supply voltage. This phenomenon results from the combined effects of shortened timing pulses at higher voltage, overcoming the increase in transistor driving strength, and larger IR drops under high parallelism. As for the DP distortion, widening the timing pulses by 1.5-2$\times$ could help prevent such RMS worsening without compromising the retention of the leaky analog data. 
 
\begin{figure}[!t]
	\centering
	\includegraphics[width=0.5\textwidth]{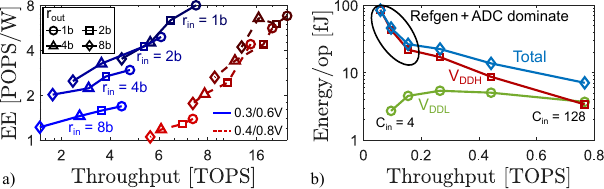}
	\caption{Trade-off between throughput and a) raw peak energy efficiency for different I/O precisions ($r_w = 1$b) and supplies under unity gain ($\mathrm{C}_{in}$ = 128), b) 8b-normalized energy/op per supply source when scaling $\mathrm{C}_{in}$.}
	\label{Fig_23}
	\vspace{-0.4cm}
\end{figure}

Thanks to the precision-configurable MBIW and SAR units, the proposed CIM-SRAM is able to trade precision for throughput with a close-to-constant energy per computing bit. Notably, Fig. \ref{Fig_23}(a) reports the measured trade-off between the macro's peak energy efficiency and throughput, evaluated at two supply voltages for different combinations of input and output precisions. As for now, binary weights and 128 input channels (i.e., 1152 rows) with unity gain are considered, while I/O transfer cycles are excluded, and weights loading through the R/W interface are neglected. In such conditions, the highest efficiency per precision value is reached for $r_{in} = r_{out} = 8$b, yielding 1.2POPS/W (i.e., 0.15POPS/W with 8b weights norm.). Having $r_{in} > r_{out}$ is unlikely as it would compress the CNN output dynamics. On the contrary, $r_{in} < r_{out}$ is common in many previous works \cite{Jia_2021, Lee_2021} to support further digital post-processing at low $r_{in}$, but degrades both throughput and efficiency. Such ADC configuration is rarely needed here thanks to the DSCI ADC architecture, as motivated in Section II. Now looking at the 8b energy breakdown in Fig. \ref{Fig_23}(b), ADCs and the resistive ladder expectedly dominate the energy/operation at low $\mathrm{C}_{in}$, as the ADC cost is merely amortized by the small-sized analog DP. As $\mathrm{C}_{in}$ increases, the disparity shrinks and both $V_{DDL}$ and $V_{DDH}$ supplies amount to a similar contribution to the total energy per operation at high $\mathrm{C}_{in}$.

\begin{table*}[!t]
\renewcommand{\arraystretch}{1.5}
\centering
\caption{Comparison to state-of-the-art CIM designs in the current, charge and digital domains}
\resizebox{\textwidth}{!}{%
\begin{threeparttable}
\begin{tabular}{@{}l|ccc|cccc|ccccc@{}}
\toprule
 & \multicolumn{3}{c|}{Digital CIM} & \multicolumn{4}{c|}{Analog current-based CIM} & \multicolumn{5}{c}{Analog charge-based CIM} \\ \midrule & \cite{Mori_2023} & \cite{Fujiwara_2022} & \cite{He_2023} & \cite{Sinangil_2021} & \cite{Yue_2022} & \cite{Houshmand_2022} & \cite{Kneip_2023} & \cite{Jia_2021} & \cite{Lee_2021} & \cite{Wang_VLSI_2022} & \multicolumn{1}{c|}{\cite{Hsieh_2023}} & \colorblue{\textbf{This work}} \\ \midrule
Year & 2023 & 2022 & 2023 & 2021 & 2022 & 2022 & 2023 & 2021 & 2021 & 2022 & \multicolumn{1}{c|}{2023} & 2023 \\
\textbf{Technology} & 
\begin{tabular}[c]{@{}c@{}}4nm\\[-0.15cm] FinFET\end{tabular} & 
\begin{tabular}[c]{@{}c@{}}28nm\\[-0.15cm] Bulk\end{tabular} & 
\begin{tabular}[c]{@{}c@{}}28nm\\[-0.15cm] Bulk\end{tabular} & 
\begin{tabular}[c]{@{}c@{}}7nm\\[-0.15cm] FinFET\end{tabular} & 
\begin{tabular}[c]{@{}c@{}}65nm \\[-0.15cm] Bulk\end{tabular} & 
\begin{tabular}[c]{@{}c@{}}22nm\\[-0.15cm] FD-SOI\end{tabular} & 
\begin{tabular}[c]{@{}c@{}}22nm \\[-0.15cm] FD-SOI\end{tabular} & 
\begin{tabular}[c]{@{}c@{}}16nm\\[-0.15cm] FinFET\end{tabular} & 
\begin{tabular}[c]{@{}c@{}}28nm\\[-0.15cm] Bulk\end{tabular} & 
\begin{tabular}[c]{@{}c@{}}22nm\\[-0.15cm] Bulk\end{tabular} & 
\multicolumn{1}{c|}{\begin{tabular}[c]{@{}c@{}}12nm\\[-0.15cm] FinFET\end{tabular}} & 
\begin{tabular}[c]{@{}c@{}}22nm\\[-0.15cm] FD-SOI\end{tabular} \\
\textbf{Bitcell type} & 2x8T + OAI & 8T & 8T-push & 8T & 8T & 12T & 6T & 8T1C & 8T1C & 8T1C & \multicolumn{1}{c|}{9T1C} & 10T1C \\
\textbf{DP mechanism} & \begin{tabular}[c]{@{}c@{}}Distributed\\[-0.15cm] adder tree\end{tabular} &
\begin{tabular}[c]{@{}c@{}}Distributed\\[-0.15cm] adder tree\end{tabular} & 
\begin{tabular}[c]{@{}c@{}}Approx.\\[-0.15cm] adder tree\end{tabular} & \begin{tabular}[c]{@{}c@{}}Capacitive\\[-0.15cm] discharge\end{tabular} &
\begin{tabular}[c]{@{}c@{}}Split-array\\[-0.15cm] cap discharge\end{tabular} & 
\begin{tabular}[c]{@{}c@{}}Capacitive\\[-0.15cm] discharge\end{tabular} &
\begin{tabular}[c]{@{}c@{}}Capacitive\\[-0.15cm] discharge\end{tabular} & \begin{tabular}[c]{@{}c@{}}Charge-inj.\\[-0.15cm] + Bit-shifting\end{tabular} &
\begin{tabular}[c]{@{}c@{}}Charge-inj.\\[-0.15cm] + Bit-shifting\end{tabular} &
\begin{tabular}[c]{@{}c@{}}C-2C\\[-0.15cm] sharing\end{tabular} & 
\multicolumn{1}{c|}{\begin{tabular}[c]{@{}c@{}}Charge inj.\\[-0.15cm] + Analog adder\end{tabular}} &
\begin{tabular}[c]{@{}c@{}}Charge inj.\\[-0.15cm] + MBIW acc.\end{tabular} \\
\textbf{On-chip CIM size} & 6.8kB & 2kB & 2kB & 16$\times$0.5kB & 2kB & 72kB & 72kB & 576kB & 36kB & 2$\times$16kB & \multicolumn{1}{c|}{16kB} & 36kB \\
\textbf{Density [kB/mm$\mathbf{^2}$]} & 395 & 40.8 & 71.4 & \colorblue{156.3} & \colordarkred{3.4} & 31.4 & \colorblue{\textbf{595.1}} & \colordarkred{23} & 70.6 & \colorblue{131} & \multicolumn{1}{c|}{-} & \colorblue{187} \\
\textbf{Supply voltage [V]} & 0.32-1.1 & 0.5-0.9 & 0.5-0.9 & 0.8-1 & 0.9 & 0.6-0.9 & 0.4/0.8 & 0.8 & 0.6 & 0.7-1 & \multicolumn{1}{c|}{0.53-0.65} & 0.3/0.6-0.4/0.8 \\
\textbf{Max precision (in/w/out)} & 16/12/36b & 4/1/4b & 8/8/8b & 4/4/4b & 8/8/8b & 7/1.5/6b & 4/4/4b & 8/8/8b & 5/1/8b & 8/8/8b & \multicolumn{1}{c|}{8/8/8b} & 8/4/8 \\
\textbf{Analog DP rescaling} & - & - & - & No & No & No & Nonlinear & No & No & No & \multicolumn{1}{c|}{No} & \colorblue{Linear} \\ \midrule
\textbf{CIM sub-system} & No & No & No & No & Yes & Yes & \colorblue{Yes} & Yes & No & No & \multicolumn{1}{c|}{No} & Yes \\
\textbf{Peak Throughput [TOPS]}\tnote{1} & N/A-0.85 & 0.01-0.325 & N/A-0.12 & 1.5-1.8 & 0.25 & \colorblue{2.2-4} & 0.08-0.4/\colorblue{0.7-4.5} & 0.25 (\colorblue{4})\tnote{2} & \colordarkred{0.1} & 0.6-1 & \multicolumn{1}{c|}{N/A-0.1} & 0.1-\colorblue{0.5} \\
\textbf{Peak Macro EE [TOPS/W]}\tnote{1} & 42.8-12.9 & 16-8 & 38-102 & \colorblue{152-109} & 19.8 & - & \colorblue{650-1050} & 51 & 91 & 32.2-15.5 & \multicolumn{1}{c|}{86.3-N/A} & \colorblue{150-125} \\
\textbf{Peak AE [TOPS/mm$\mathbf{^2}$]}\tnote{1} & N/A-49.9 & 0.16-6.3 & 4.2 & \colorblue{23-36} & 0.04 & \colorblue{1.6-2.1\tnote{3}} & \colorblue{0.27-1.6\tnote{2} / 6-36\tnote{3}} & 0.7 & 0.18 & 2.4-4 & \multicolumn{1}{c|}{-} & 0.51-\colorblue{2.6} \\
\textbf{Peak System EE [TOPS/W]}\tnote{1}\tnote{1} & - & - & - & - & 4.4 & \colorblue{49-29} & \colorblue{112-204} & 30 & - & - & \multicolumn{1}{c|}{-} & \colorblue{40-35} \\
\textbf{Max 8b output RMS [LSB]} & - & - & - & - & - & \colordarkred{0.8-2} & \colordarkred{1.7-2.15} & 0.9 & 0.98 & 0.1 & \multicolumn{1}{c|}{0.3} & 0.32-1.8 \\
\textbf{MNIST acc. [\%]} & - & - & - & 99.6 & 99.3 & - & 98.1\tnote{5} & - & - & 98.1 & \multicolumn{1}{c|}{-} & 98.6\tnote{4} \\
\textbf{CIFAR-10 acc. [\%]} & - & 90.4 & 91.6 & 88.5 & 92.9 & 89.3 (mixed) & \colordarkred{$\sim$ 73}\tnote{5} & 91.5 & 91.1 & - & \multicolumn{1}{c|}{-} & 90.8\tnote{5} \\ \bottomrule
\end{tabular}
\begin{tablenotes}
\item[1] Normalized to 8b precision (inputs \textit{and} weights).
\item[2] Includes I/O transfer cycles.
\item[3] Dependent on timings config.
\item[4] Measured on chip for a modified 4b LeNet-5.
\item[5] Emulated post-silicon on a 4b VGG-16.
\end{tablenotes}
\end{threeparttable}
}
\vspace{-0.2cm}
\label{Table1}
\end{table*}

\subsection{CIM-CNN Accelerator Breakdown}
\begin{figure}[!t]
	\centering
	\includegraphics[width=0.5\textwidth]{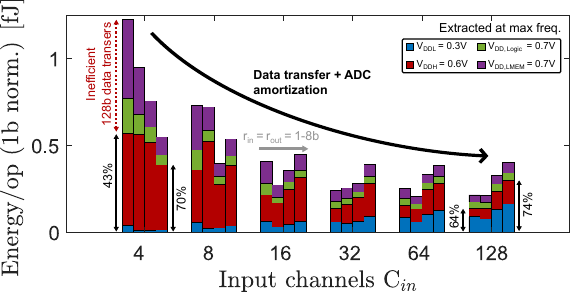}
	\caption{Energy breakdown of the CIM-CNN accelerator in convolutional mode with increasing input channel and precision configurations (unity gain), at the 0.3/0.6V maximum frequency per point (unity gain $\gamma$).}
	\label{Fig_24}
	\vspace{-0.4cm}
\end{figure}

The overall throughput and efficiency of the whole CIM-CNN accelerator is altered by I/O data transfers to/from the macro. In convolutional mode, the standalone CIM-SRAM throughput is divided by the number of transfer cycles from Eq. (\ref{Eq_N_in}) or (\ref{Eq_N_out}), where we consider the same clock frequency for the macro and its digital datapath. As such, Fig. \ref{Fig_24} gives the channel- and precision-wise evolution of the CIM-CNN energy per operation, normalized to 1b and extracted at each configuration's maximum operating frequency for a 0.3/0.6V supply. Another dedicated test mode ensures accurate power estimates by looping on the convolution operation of a 32$\times$32 image, with randomly distributed inputs and weights. Now, the energy per operation decreases with $\mathrm{C}_{in}$ thanks to a better amortization of the ADC overhead (including the resistive ladder DC power) and data transfers. In particular, layer configurations using less than 128b have their total energy dominated by data transfers, which cannot fully make use of the available system bandwidth. On the contrary, the macro's energy accounts for 70-75\% of the total accelerator energy in high precision and/or channel configuration. However, it becomes sensitive to leakage integrated over the high number of I/O transfers in the MHz-range. To prevent this while further boosting throughput, combining a higher clock frequency for pipelined transfers with multi-cycle CIM-SRAM operations, as described in Section IV, could be considered. Such solution was however not implemented here due to the higher inbound complexity of the CIM-SRAM's internal time generator as well as design-time constraints, but should be considered in future solutions. 

\subsection{Comparison to the State of the Art}
The proposed macro and accelerator are compared to other state-of-the-art CIM architectures in Table \ref{Table1}, including recent designs in the current, charge and digital domains. IMAGINE achieves the highest density thanks to its custom bitcell layout and area-optimized SAR topology, while reaching a 3-to-5$\times$ improved macro-level peak energy efficiency across charge-domain architectures. It also achieves a competitive 0.5TOPS throughput and a 40TOPS/W peak accelerator efficiency while being the first work to provide linear in-memory offset and gain rescaling abilities, therefore skipping the need for most inter-layer processing during CNN execution (whose overhead is usually not reported). Although the accelerator's throughput and energy efficiency are lower than current-based designs, these suffer from a much higher variability on their digitized result, which jeopardizes the mapping of applications requiring a medium computing precision. Compared to previous charge-based designs, the serial-split DPL used in IMAGINE leads to a slightly smaller throughput and higher RMS following the insufficient timing configuration in the SS corner. However, its compact ADC leads to a higher energy efficiency, while the DPL adaptivity brings further gains in non-peak utilization scenarios. Finally, compared to the nonlinear charge-integrating ABN in  \cite{Kneip_2023}, the linear approach in this work supports more realistic workloads than MNIST, e.g., CIFAR-10 and beyond. Altogether, this work extends the boundaries of the computing efficiency versus accuracy of moderate-precision CIM designs, achieving the best charge-based efficiency to date while supporting flexible workloads thanks to its adaptive dynamic range utilization. 

\section{Conclusion}
In this work, we presented IMAGINE, a 1-to-8b compute-in-memory (CIM) CNN accelerator embedding a charge-based CIM-SRAM with a multi-bit input-serial, weight-parallel (MBIW) end-to-end dot-product (DP) operation. Namely, a split DP line structure allows up to 20$\times$ higher voltage swing utilization, depending on the input channel configuration, while in-ADC 5b offset and rescaling effects enable linear data shaping, making full use of the selected ADC precision. Standalone measurement results of the dense 187kB/mm$^2$ macro showcase peak energy and area efficiencies of 0.15-to-8POPS/W and 2.6-154TOPS/mm$^2$, respectively, with quasi-linear scaling from 8 to 1b computing, exceeding previous charge-based CIM-SRAM designs by up to 5$\times$. Moreover, the mean measured 8b RMS error under unity gain lays below one LSB, similar to previous works. Still, in-ADC scaling also upscales the RMS error, bringing computing LSBs below the macro's noise floor for 8b computations. Although noise-aware CNN training partly deals with this loss on medium complexity tasks, further mismatch and noise reduction would be required for 8b applications targeting gain values above 16, at the cost of additional energy and area. At the system level, the entire CIM-CNN accelerator reaches a 40TOPS/W peak energy efficiency and an overall throughput comparable to previous works. Nonetheless, it could be further improved by enabling multi-cycle macro operation or, possibly, a precision-last dataflow based on intertwined analog operations and digital transfers. Such approach could bring the intrinsically slower charge-based designs closer to the throughput and system-level efficiency of current-based CIM-SRAM approaches, while taking advantage of their much lower variability.

\appendices

\section*{Acknowledgment}
The authors would like to thank ECS team members for their help with the CERBERUS chip design and proofreading of this work, in particular M. Gonzalez and S. Favresse.

\ifCLASSOPTIONcaptionsoff
  \newpage
\fi


\bibliographystyle{IEEEtran}
\bibliography{Sections/biblio}

\end{document}